\documentclass[12pt]{article}
\usepackage{graphicx,amssymb,amsfonts}
\newlength{\extraspace}
\newlength{\extraspaces}
\newcommand{\be}{\begin{equation}
\addtolength{\abovedisplayskip}{\extraspaces}
\addtolength{\belowdisplayskip}{\extraspaces}
\addtolength{\abovedisplayshortskip}{\extraspace}
\addtolength{\belowdisplayshortskip}{\extraspace}}
\newcommand{\ee}{\end{equation}}

\newcommand{\ba}{\begin{eqnarray}
\addtolength{\abovedisplayskip}{\extraspaces}
\addtolength{\belowdisplayskip}{\extraspaces}
\addtolength{\abovedisplayshortskip}{\extraspace}
\addtolength{\belowdisplayshortskip}{\extraspace}}
\newcommand{\ea}{\end{eqnarray}}

\newcommand{\nonu}{\nonumber \\[.5mm]}
\newcommand{\A}{&\!\!\!}

\newcommand{\newsection}[1]{
\vspace{7mm} \pagebreak[3] \addtocounter{section}{1}
\setcounter{subsection}{0} \setcounter{footnote}{0}
\begin{center}
{\large {\bf \thesection. #1}}
\end{center}
\nopagebreak
\medskip
\nopagebreak \hspace{3mm}}

\begin{document}

\begin{center}
{{\bf  Cosmological applications in Kaluza-Klein theory}}
\end{center}
\centerline{ M. I. Wanas${}^{a,d,e}$, Gamal G.L.
Nashed${}^{b,c,d,e}$ and A. A. Nowaya${}^{f}$}

\bigskip

\centerline{\it Astronomy Department, Faculty of Science, Cairo
University, Giza, Egypt ${}^a$}\vspace*{0.4cm} \centerline{\it
Mathematics Department, Faculty of Science, King Faisal
University, P.O. Box.} \centerline{\it 380 Al-Ahsaa 31982, the
Kingdom of Saudi Arabia
 ${}^b$\footnote{ Mathematics Department, Faculty of Science, Ain
Shams University, Cairo, 11566, Egypt ${}^c$ \vspace*{0.4cm}\\
Center for Theoretical Physics, British University of  Egypt
 Sherouk City 11837, P.O. Box 43, Egypt ${}^d$
\vspace*{0.4cm} \\
 Egyptian Relativity Group (ERG) URL:
http://www.erg.eg.net ${}^e $}} \vspace*{0.4cm} \centerline{\it
Physics Department, Faculty of Science, Monofia University,
Monofia, Egypt ${}^f $} \vspace*{0.4cm}

\bigskip
\hspace{2cm}
\\
\\
\\
\\
\\
\\

\abstract{
The field equations of Kaluza-Klein (KK) theory have
been applied in the domain of cosmology. These  equations are
solved for a flat universe by taking the gravitational and  the
cosmological constants as a function of time t. We use  Taylor's
expansion of cosmological function, $\Lambda(t)$, up to the first
order of the time $t$.  The cosmological parameters  are
calculated and some cosmological problems are discussed.}
\vspace{.2cm}\\
\hspace*{.1cm}{\it Keywords: Kaluza-Klein  theory, cosmology,
Taylor's expansion of cosmological function }\vspace{0.01cm}\\\\
\hspace*{0.1cm}{\it PACS numbers:0455, 0490, 9880D  }

\begin{center}
\newsection{\bf Introduction}
\end{center}

Recent  observation of  distant supernovae type Ia indicate that
the universe is presently accelerating$^{ \rm [1-4]}$. The cosmic
acceleration is attributed to the presence of an unknown form of
energy violating the strong energy conditions $\rho  +3p > 0$
where $\rho $ and $p$ are energy density and pressure of dark
energy, respectively. Different candidates, for dark energy$^{ \rm
[5,6]}$, are attempted to yield accelerating cosmologies at late
time . The cosmological constant $\Lambda$ and phantom fields$^{
\rm [7-19]}$ violating a weak energy conditions $\rho +p > 0$ are
most popular ones.

Kaluza's$^{ \rm [20]}$ achievement has been to show that
five-dimensional general relativity contains both Einstein's
4-dimensional theory of gravity and Maxwell's theory
 of electromagnetism. He however imposed a somewhat artificial restriction
(the cylindrical condition) on the coordinates, essentially
barring the fifth one a priori from making a direct appearance in
the laws of physics. Klein's$^{ \rm [21]}$ contribution
 was to make this restriction less artificial by suggesting a plausible
physical basis for it in compactification of the fifth dimension
(for more details see Ref. \cite{OW8}). This idea was
enthusiastically received by unified-field theorists, and when the
time came to include the strong and weak forces by extending
Kaluza's mechanism to higher dimensions, it was assumed that these
too would be compact. This line of thinking has led through
eleven-dimensional supergravity theories in the 1980 to the
current favorite contenders for a possible "theory of everything,"
ten-dimensional superstrings.

Most existing astrophysical  work on the cosmological term,
operate on the assumption that $\Lambda$ is a constant. However,
quantum field theorists and others are treating the cosmological
term as a dynamical quantity
  (cf. \cite{Bp}$\sim$\cite{Pa1}). Anything which contributes to the energy density $\rho$ of the vacuum behaves like a cosmological
term via $\Lambda=8\Pi G \rho$. Many potential sources of
fluctuating vacuum energy have now been identified, including
scalar fields$^{ \rm [31-39]}$, tensor fields$^{ \rm [40-44]}$,
nonlocal effects$^{ \rm [45,46]}$, wormholes$^{ \rm [47,48]}$,
inflationary mechanisms$^{ \rm [49]}$ and even cosmological
perturbations.$^{ \rm [50]}$ Each of these has been claimed to
give rise to a negative energy density which grows with time,
tending to cancel out any pre-existing positive cosmological term
and drive the net value of $\Lambda$ toward zero. Processes of
this kind are among the most promising ways to resolve the
longstanding cosmological "constant" problem$^{ \rm [31]}$ (see
\cite{Ws9} for review).

 It is the
aim of the present work to solve the   gravitational  field
equations of Kaluza-Klein (KK) theory in the domain of cosmology
considering the gravitational and cosmological constants  as
function of time $t$.
 Then, we assume that $\Lambda(t)=\epsilon R(t)^{-2}$, where $R(t)$ is the scale factor used to calculate the cosmological parameters. In  \S 2,
 the field equations of KK are applied to the FRW metric.  In \S 3,
we give  solution of the field equation  and calculate  the
corresponding cosmological parameters. In \S 4, we give the
solution of the field equation for a closed universe and also
calculate  the cosmological parameters associated with this model.
The final section is devoted to discussion and conclusion.
\newpage
\newsection{The FRW model and  KK theory }

The FRW line element  in 5-dim has the form$^{ \rm [52]}$ \be
ds^2=dt^2-R^2(t)\Biggl[\frac{dr^2}{1-kr^2}+r^2\left(d\theta^2+\sin^2\theta
d\phi^2\right)+(1-kr^2)d\psi^2\Biggr], \ee where $R(t)$ is the
scale factor, $k$ is the curvature parameter ($k=+1,0,-1$)
corresponding to closed flat and open universe respectively. The
Einstein field equations  are given by \be R_{\mu \nu}-g_{\mu
\nu}\frac{R}{2}=g_{\mu \nu} \Lambda-8\pi GT_{\mu \nu}=8\pi
GT'_{\mu \nu}, \ee where $R_{\mu \nu}$, $R$, $g_{\mu \nu}$,
$\Lambda=\Lambda(t)$, $G=G(t)$ are Ricci tensor, Ricci scalar,
metric tensor, cosmological function and gravitational function,
respectively.

 Assuming that matter filling the universe is in the form of a perfect fluid, then
\be {T'_0}^0=\rho+\frac{ \Lambda}{8\pi G}, \qquad \qquad
{T'_1}^1={T'_2}^2={T'_3}^3={T'_4}^4=-P+\frac{ \Lambda}{8\pi G},\ee
where $\rho=\rho(t)$ is the density and $P=P(t)$ is the pressure.
Applying Eq. (2) to metric given by Eq. (1) and using Eq. (3)  we
get the following set of differential equations \be 8\pi G
\rho+\Lambda = 6\Biggl[\left(\frac{\dot
R}{R}\right)^2+\frac{k}{R^2}\Biggr],\ee \be
 8\pi GP-\Lambda = -3\Biggl[\frac{\ddot R}{R}+\left(\frac{\dot {R}}{R}\right)^2+\frac{k}{R^2}\Biggr].\ee
  Now we have two differential equations in five unknown functions
 $R(t)$, $\rho(t)$, $P(t)$, $\Lambda(t)$ and $G(t)$. Therefore we need some extra
 conditions to be able to solve Eqs. (4) and (5), for each value of
 k. One of these  conditions is usually provided by an equation of
 state  characterizing the fluid filling the model. Here we take  the equation of
 state  in the form
\be P=(\omega-1)\rho,\ee where  $\omega \in [0,2]$. When
$\omega=0$, then  $P=-\rho$ (dark energy case),  when $\omega=1$,
then  $P=0$ (dust case),  when $\omega= \frac{4}{3}$ then
$P=\frac{1}{3}\rho$ (radiation case) and when $\omega= 2$ then
$P=\rho$ (stiff matter case). The conservation  equation has the
form ${T'^{\mu \nu}}_{; \ \nu}=0.$ This will gives rise to the
relation,\be \dot{\rho}+4\frac{\dot R(t)}{R(t)} \left(\rho+P
\right)+\frac{\Lambda}{8\pi
G}\left(\frac{\dot{\Lambda}}{\Lambda}-\frac{\dot{G}}{G}\right)=0.\ee
Eqs (4), (5) and (6) are three nonlinear differential equations in
the five unknown functions. Therefore,  we spilt Eq. (7), for
simplicity, into the two equations \be
 \dot{\rho}+4\frac{\dot R(t)}{R(t)} \left(\rho+P \right)=0,\ee
 and
\be \frac{\dot{\Lambda}}{\Lambda}-\frac{\dot{G}}{G}=0,\ee in order
to impose two further additional conditions on the unknown
function. Now, we have five differential equations (4), (5), (6),
(8) and (9) in the five unknown functions $R(t)$, $\rho(t)$,
$P(t)$, $\Lambda(t)$ and $G(t)$.

 Eq. (9) shows that $\Lambda$ is constant whenever $G$ is constant and vice versa. Solving Eq. (8), we obtain
\be \rho(t)=\frac{c_1}{R(t)^{4\omega}}, \ee where Eq. (6) has been
used and $c_1$ is a constant of integration. The constant $c_1$
can be determined by using $\omega=\omega_0$ and $\rho=\rho_c$
(critical density) at time $t=t_0$, where the subscript $0$ refers
to the present value. We are going to use the following
definitions of some cosmological parameters. \ba \A \A The \ \
Hubble's \ \ parameter \ \ H_0={\dot{R(t_0)} \over
R(t_0)}={\dot{R}_0 \over R_0},\nonu
 \A \A the \ \ Hubble's \ \ time \ \  \tau_0 \stackrel{\rm def.}{=} {H^{-1}}_0, \nonu
 \A \A the \ \
deceleration \ \ parameter \ \  q_0 \stackrel{\rm def.}{=} -
{\ddot{R}_0 \over R_0}{H^{-2}}_0=-1-\frac{\dot{H_0}}{{H_0}^2},
\nonu
\A \A the \ \ density \ \ parameter \ \ \sigma_0 \stackrel{\rm
def.}{=}  {\rho_0 \over \rho_c}, \nonu
\A \A where \ \ \rho_c \ \ is \ \ the \ \ critical \ \ density \ \
defined \ \ by \ \
 \rho_c \stackrel{\rm def.}{=} {3 {H^{2}}_0 \over 8\pi G}. \ea
 \newpage
 \newsection{Solution for k=0 }
 We are going to solve the field equations\footnote {In Ref. \cite{Sm1} a solution of the field equations (4), (5) has been obtained for the case k=0,  $\Lambda(t) \simeq \epsilon H^{2}(t)$ and the relevant cosmological parameters are also calculated. The continuity equation given in this case, is the same as Eq. (8),  and not like Eq. (7), which does not include the variation of $ \Lambda(t)$ and $G(t)$.}  (4), (5) and then we evaluate the cosmological parameters.  Here we assume\footnote{In these calculations we will take Taylor's expansion, of $\Lambda(t)$, up to first order in $(t)$ and we will neglect higher order.}   \be \Lambda(t) \simeq \epsilon_0+\epsilon_1t+\cdots,  \ee where $\epsilon_0$ and  $\epsilon_1 \cdots$ are arbitrary constants. From Eq. (9) we get \be G(t)=c_2( \epsilon_0+\epsilon_1t),\ee where $c_2$ is another constant of integration.

 Eqs. (4) and (5), in the case $k=0$, take the forms
 \ba 8\pi G \rho+\Lambda\A =\A 6\Biggl[\left(\frac{\dot R(t)}{R(t)}\right)^2\Biggr],\nonu
 8\pi GP-\Lambda\A =\A -3\Biggl[\frac{\ddot R(t)}{R(t)}+\left(\frac{\dot {R(t)}}{R(t)}\right)^2\Biggr].\ea
 From Eq. (14)  we get
 \be 3\frac{\ddot R(t)}{R(t)}-3\left(\frac{\dot {R}(t)}{R(t)}\right)^2+\omega\Biggl[6\frac{\dot {R^2}(t)}{R^2(t)}-\Lambda\Biggr]\equiv\dot{H(t)}+\omega\Biggl[2H^2(t)-\frac{\epsilon_0}{3}-\frac{\epsilon_1t}{3}\Biggr]=0,\ee where   Eq. (6) has been used again.
  Eq. (15) is a differential equation in two variables so we are going to study the following cases, dust, radiation and stiff matter.

  We will discard the case $\omega=0$ since this will makes the density parameter, given by Eq. (10), always constant, in spite of the
  expansion of the universe. This needs  an extensive discussion which will be done in the future work.

 \underline{Dust case ( $\omega= 1)$}:\vspace{0.2cm}\\ The integration of Eq. (15) has the form
 \ba H(t) \A=\A A_0\frac{c_3 A\left(1,\frac{\Lambda(t)}{A_{00}}\right)-c_4B\left(1,\frac{\Lambda(t)}{A_{00}}\right)}{c_3 A\left(\frac{\Lambda(t)}{A_{00}}\right)-c_4B\left(\frac{\Lambda(t)}{A_{00}}\right)}, \nonu
  A_0 \A=\A \left(\frac{\epsilon_1}{12}\right)^{1/3}, \quad A_{00}=\left(\frac{3{\epsilon_1}^2}{2}\right)^{1/3}, \qquad  c_3\ \ and \ \  c_4
   \ \ are  \ \   constants \ \ of \ \  integration,\nonu
   \ea where
 $A(1,x)$ and  $B(1,x)$ are the AiryAi(1,x) and AiryBi(1,x) wave functions\footnote{AiryAi and  AiryBi(z) are linearly independent solutions for w in the equation $w''-zw=0$. Specifically $AiryAi(z)=C_1 F_1(2/3;z^3/9)-C_2F_1(4/3;z^3/9)$ and $AiryBi(z)=\sqrt{3}\left[{C^\ast}_1 F_1(2/3;z^3/9)+{C^\ast}_2F_1(4/3;z^3/9)\right]$ where ${C^\ast}_1=AiryAi(0)$ and ${C^\ast}_2=AiryAi'(0)$ and $F_1$ is the generalized hypergeometric function. Also $AiryAi(1,x)$ is the 1st. derivative of $AiryAi(x)$.}.

  From Eq. (16)  we get the scale factor in the form

 \be R(t)=\sqrt{\frac{c_3 A\left(\frac{\Lambda(t)}{A_{00}}\right)-c_4B\left(\frac{\Lambda(t)}{A_{00}}\right)}{A_0\left[A\left(1,\frac{\Lambda(t)}{A_{00}}\right)
 B\left(\frac{\Lambda(t)}{A_{00}}\right)-B\left(1,\frac{\Lambda(t)}{A_{00}}\right)
 A\left(\frac{\Lambda(t)}{A_{00}}\right)\right]}}.\ee

 \underline{Radiation case ( $\omega= \frac{4}{3})$}:\vspace{0.2cm}\\ The integration of Eq. (15) has the form
 \ba H(t) \A=\A B_0\frac{c_5 A\left(1,\frac{\Lambda(t)}{B_{00}}\right)c_6 B\left(1,\frac{\Lambda(t)}{B_{00}}\right)}{c_5 A\left(\frac{\Lambda(t)}{B_{00}}\right)-c_6 B\left(\frac{\Lambda(t)}{B_{00}}\right)}, \nonu
  B_0 \A=\A \left(\frac{{\epsilon_1}^3}{16}\right)^{1/3}, \quad B_{00}=\left(\frac{27{\epsilon_1}^2}{32}\right)^{1/3},\qquad  c_5\ \ and
  \ \  c_6 \ \ are  \ \   constants \ \ of \ \  integration.\nonu
  \A \A \ea
   From Eq. (18)  we get the scale factor in the form
 \be R(t)=\left\{ \frac{c_5 A\left(\frac{\Lambda(t)}{B_{00}}\right)-c_6B\left(\frac{\Lambda(t)}{B_{00}}\right)}{B_0\left[A\left(1,\frac{\Lambda(t)}{B_{00}}\right)
 B\left(\frac{\Lambda(t)}{B_{00}}\right)-B\left(1,\frac{\Lambda(t)}{B_{00}}\right)
 A\left(\frac{\Lambda(t)}{B_{00}}\right)\right]}\right\}^{\frac{3}{8}}.\ee

\underline{Stiff matter case ( $\omega= 2)$}:\vspace{0.2cm}\\ The
integration of Eq. (15) has the form
 \ba H(t)\A =\A C_0\frac{c_7 A\left(1,\frac{\Lambda(t)}{C_{00}}\right)-c_8 B\left(1,\frac{\Lambda(t)}{C_{00}}\right)}{c_5 A\left(\frac{\Lambda(t)}{C_{00}}\right)+B\left(\frac{\Lambda(t)}{C_{00}}\right)}, \nonu
 \A \A C_0=\left(\frac{\epsilon_1}{24}\right)^{1/3}, \quad C_{00}=\left(\frac{3{\epsilon_1}^2}{8}\right)^{1/3}, \qquad  c_7\ \ and \ \
 c_8 \ \ are  \ \   constants \ \ of \ \  integration.\nonu
   \A \A \ea  From Eq. (20)  we get the scale factor in the form
 \be R(t)=\left\{\frac{c_7 A\left(\frac{\Lambda(t)}{C_{00}}\right)-c_8 B\left(\frac{\Lambda(t)}{C_{00}}\right)}{C_{00}\left[A\left(1,\frac{\Lambda(t)}{C_{00}}\right)
 B\left(\frac{\Lambda(t)}{C_{00}}\right)-B\left(1,\frac{\Lambda(t)}{C_{00}}\right)
 A\left(\frac{\Lambda(t)}{C_{00}}\right)\right]}\right\}^{1/4}.\ee
 \begin{figure}
\begin{center}
{\includegraphics[width=16cm]{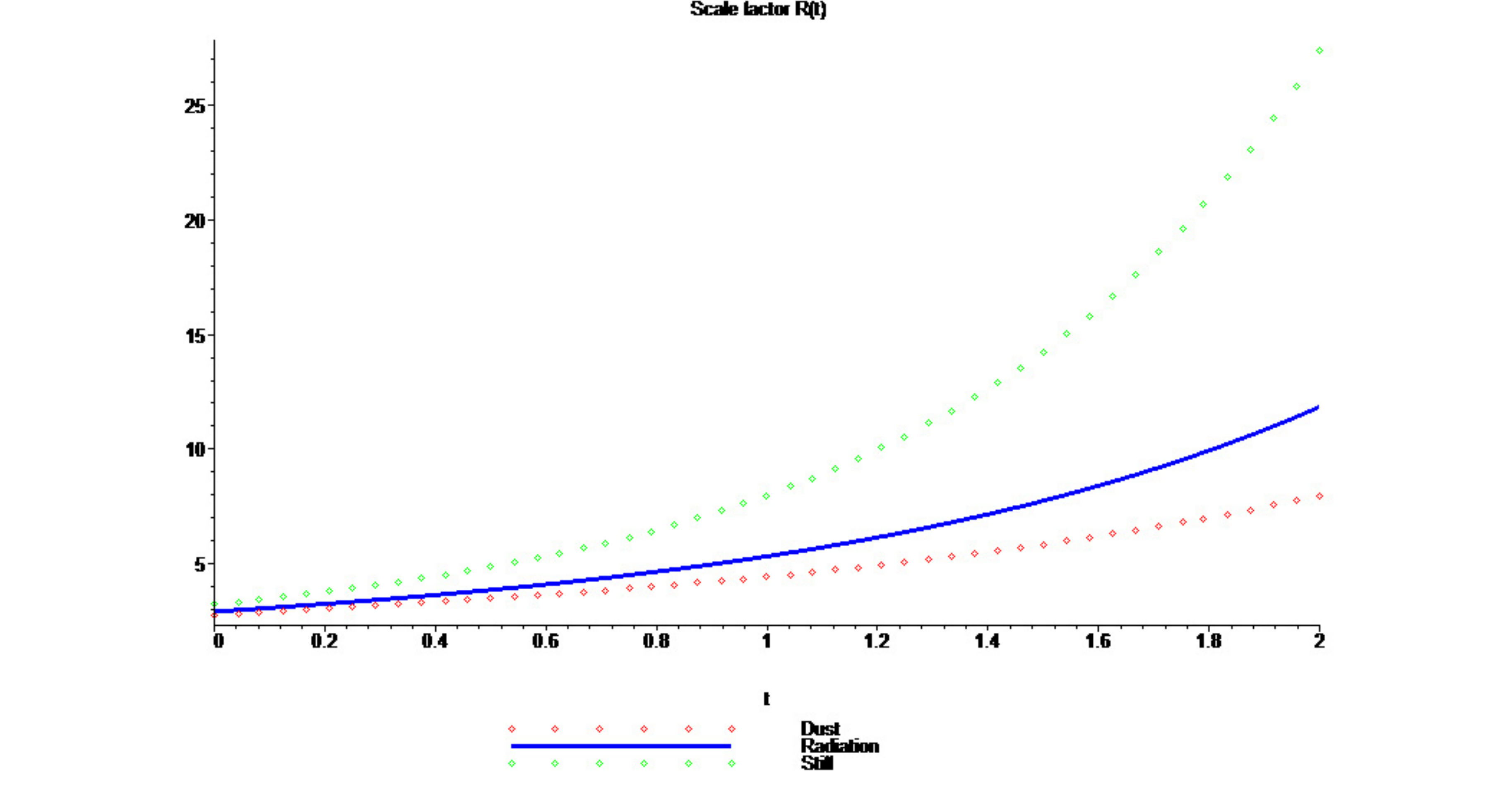}}
 \caption{The Scale factor $R(t)$ of different cases  for $\epsilon_0=1$, $\epsilon_1=1$, $c_3=1$, $c_4=1$ , $c_5=1$ , $c_6=1$ , $c_7=1$ , $c_8=1$.}
\end{center}
\begin{center}
{\includegraphics[width=13cm]{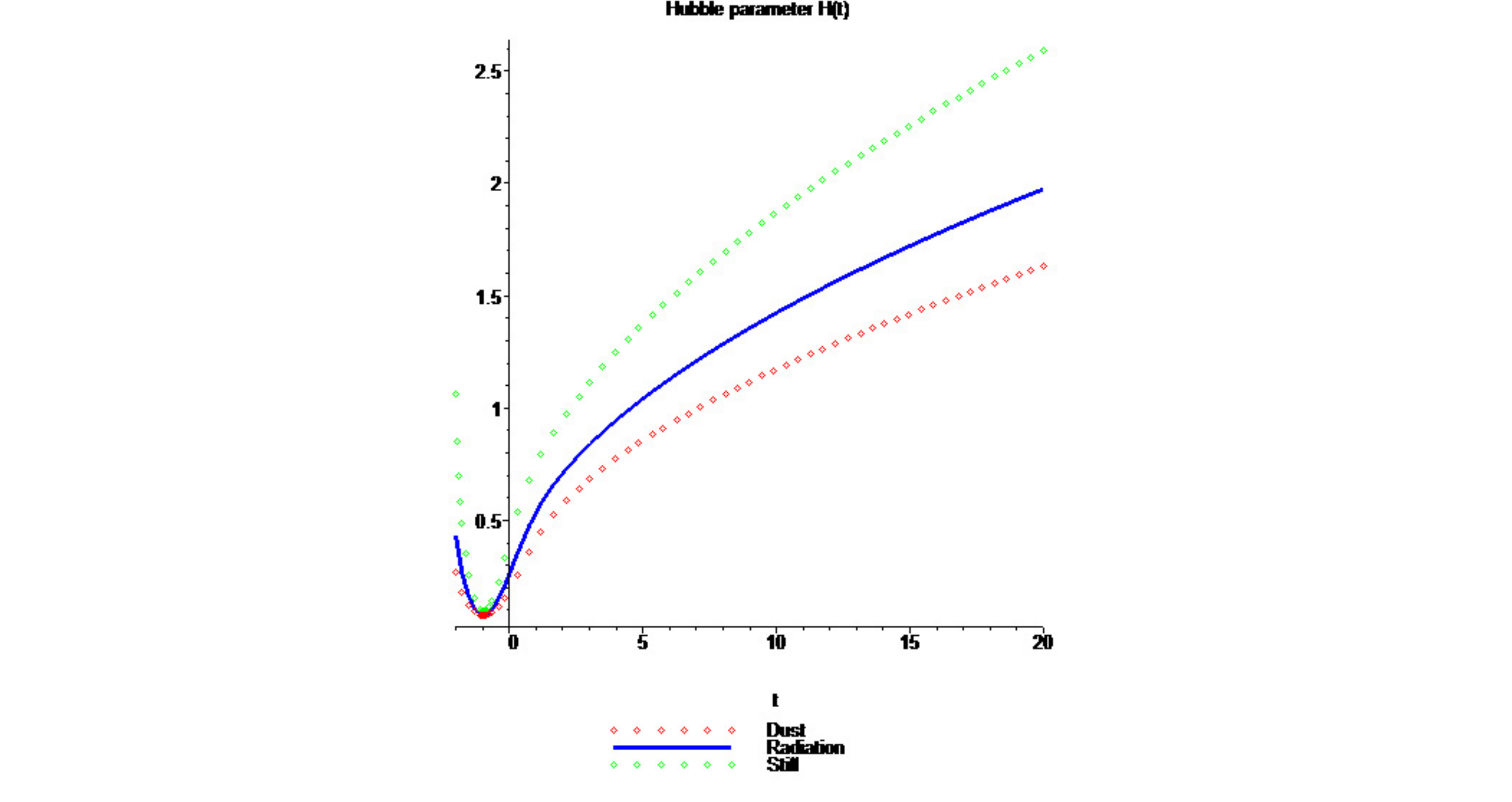}}
 \caption{The Hubble parameter $H(t)$ of different cases  for $\epsilon_0=1$, $\epsilon_1=1$, $c_3=1$, $c_4=1$ , $c_5=1$ , $c_6=1$ , $c_7=1$ , $c_8=1$.}
\end{center}
\end{figure}
 \begin{figure}
\begin{center}
{\includegraphics[width=11cm]{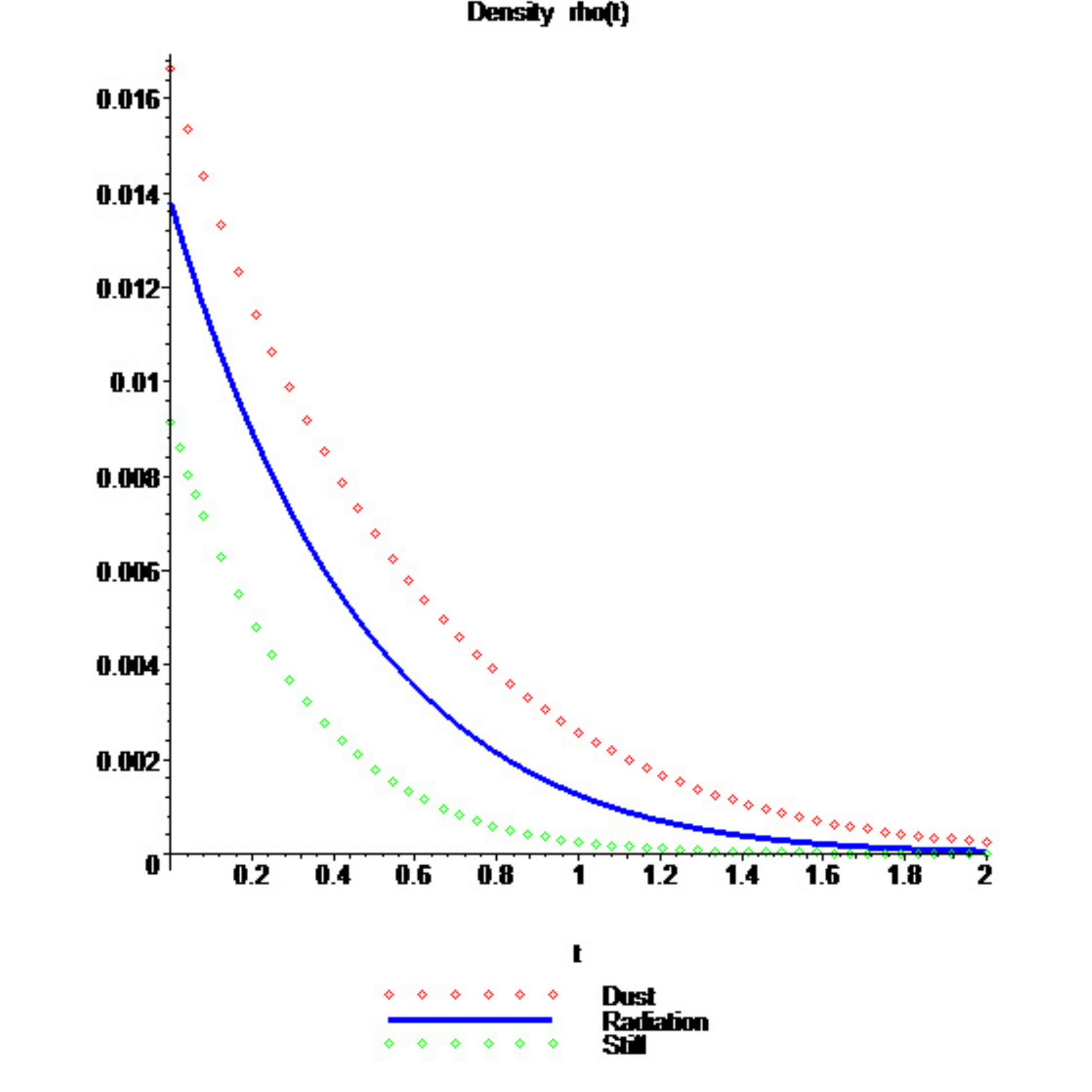}}
 \caption{The density  $\rho(t)$ of different cases  for $\epsilon_0=1$, $\epsilon_1=1$, $c_3=1$, $c_4=1$ , $c_5=1$ , $c_6=1$ , $c_7=1$ , $c_8=1$.}
\end{center}
\begin{center}
{\includegraphics[width=11cm]{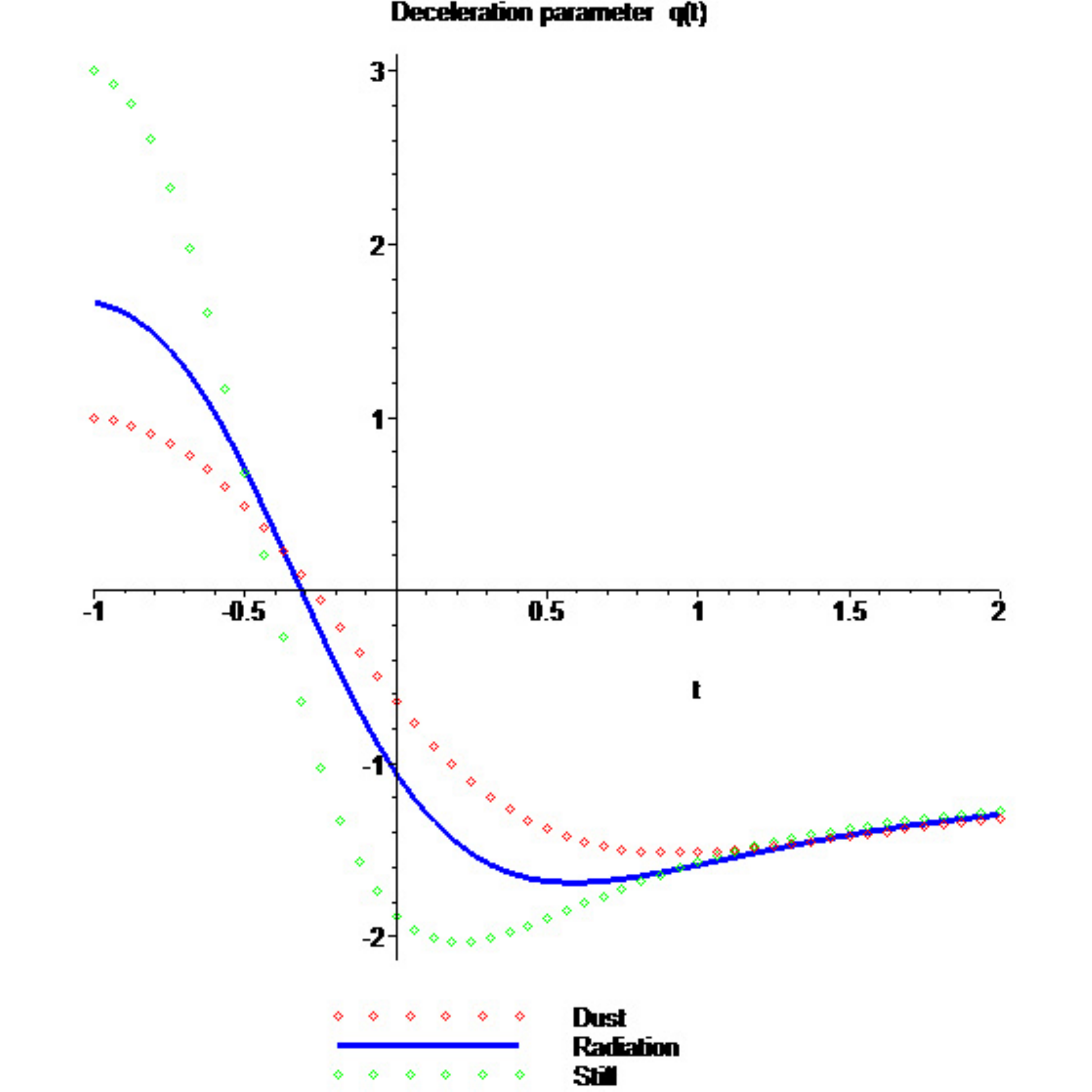}}
 \caption{The deceleration parameter $q(t)$  of different cases  for $\epsilon_0=1$, $\epsilon_1=1$, $c_3=1$, $c_4=1$ , $c_5=1$ , $c_6=1$ , $c_7=1$ , $c_8=1$.}
\end{center}
\end{figure}
\begin{figure}
\begin{center}
{\includegraphics[width=9cm]{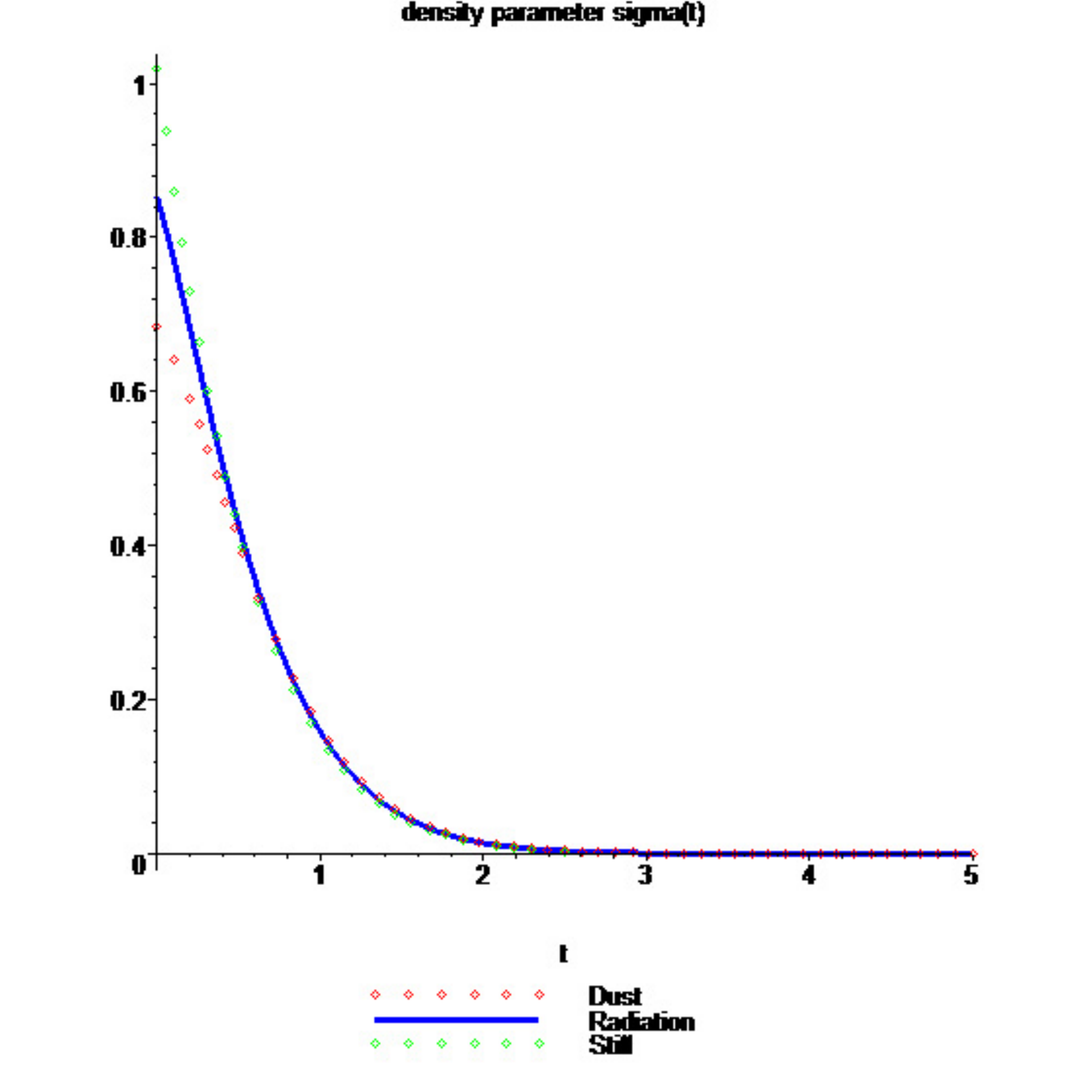}}
 \caption{The density parameter $\sigma(t)$ of different cases  for $\epsilon_0=1$, $\epsilon_1=1$, $c_3=1$, $c_4=1$ , $c_5=1$ , $c_6=1$ , $c_7=1$ , $c_8=1$.}
\end{center}
\end{figure}
\newsection{Physical properties of the  models obtained}
 \underline{\bf Red shift}:
 Consider an observer O, located at the origin of space
 coordinate and a galaxy S at an arbitrary point with spatial
 coordinates $(r, \ \theta, \ \phi, \  \psi)$. According to Eq. (1), the
 velocity of a radial ray of light $(ds=0, \ d\theta=0, \
 d\phi=0, \  d\psi=0)$ is given by \be \displaystyle{dr \over dt}=\pm {\sqrt{1-kr^2}
 \over R(t)}.\ee Hence if a light pulse is emitted by the galaxy S
 at the instant $t_1$, is received by O at the instant $t_0$,
 $(t_0>t_1)$ we get from (22) \be f(r_1)=- {\int_{t_1}}^{t_0} \displaystyle{dt \over R(t)}=\left \{
\begin{array}{ll} \sin^{-1} r_1 \qquad \qquad k=+1
, \vspace{.2cm}\\ r_1 \qquad \qquad \qquad \quad k=0,  \vspace{.2cm}\\
\sinh^{-1} r_1 \qquad \qquad k=-1.
\end{array} \right. \ee
 The value of
the red-shift Z in all obtained models can be proved to has the
form \be Z={\delta \lambda
\over \lambda}={R_0 \over R_1}-1.\ee \vspace{.2cm}\\
 \underline{Particle Horizons}:\\

 One  of the
important questions in cosmological theory is to find out,
according to a certain model, the answer of the following
questions: \\ How much of our universe can be observed at a given
instant? To answer this question we consider, as above a galaxy S
at the point $(r, \theta, \phi, \  \psi)$ and an observer O,
located at the origin space coordinates. Then consider a light
signal emitted by S at the instant $t_1$ and received by O at the
instant $t_0$. Light signal moves along null geodesic whose
equation gives rise to \be
\frac{dr}{dt}=\frac{\sqrt{1-kr^2}}{R(t)} \Rightarrow
\frac{dr}{\sqrt{1-kr^2}}=\frac{dt}{R(t)}=f(r_1),\ee

We have to distinguish between two different distances, the radial
coordinate distance defined by \be r \stackrel{\rm def.}{=}
{\int_{t_1}}^{t_0} \displaystyle{dt \over R(t)}, \ee and the
proper distance $s$ defined by \be s  \stackrel{\rm def.}{=}
R(t_0) r. \ee Recalling the red-shift as given by (24), one can
always express the proper distance $s$ in terms of the red-shift
$Z$. Then if $s \rightarrow \infty$ as $Z\rightarrow \infty$, we
can say that the model is free from particle horizons, i.e.
isotropic observers can communicate by sending and receiving
signals. But if $s$ attains a finite value as $Z\rightarrow
\infty$ we say that the model contains particle horizons.

For example we are going to explore the existence of particle
horizons in one of the models discussed above. For the dust model
$(\omega=1, k=0, \Lambda(t)=\epsilon_0+\epsilon_1t)$,  given by
(17), we have:
\[f(r_1)=r_1= {\int_{t_1}}^{t_0} \displaystyle{dt \over
R(t)}={\int_{t_1}}^{t_0}\displaystyle {dt \over \sqrt{\frac{c_3
A\left(\frac{\Lambda(t)}{A_{00}}\right)-c_4B\left(\frac{\Lambda(t)}{A_{00}}\right)}{A_0\left[A\left(1,\frac{\Lambda(t)}{A_{00}}\right)
 B\left(\frac{\Lambda(t)}{A_{00}}\right)-B\left(1,\frac{\Lambda(t)}{A_{00}}\right)
 A\left(\frac{\Lambda(t)}{A_{00}}\right)\right]}}} \]
 Taking  Taylor's expansion of the above equation, up to first order in $t$, we get
\[r_1={\int_{t_1}}^{t_0}\displaystyle {dt \over
\sqrt{-\displaystyle\frac{c_3 \Gamma[\frac{1}{3}]}{2 \sqrt[6]{3}}
+ \frac{ c_4\sqrt[3]{3}\Gamma[\frac{1}{3}]}{2} } +
\displaystyle\frac{
 t \left(\sqrt{2} \sqrt[3]{9} c_3 \Gamma[\frac{2}{3}] +
    3 \sqrt{2} \sqrt[6]{3} c_4 \Gamma[\frac{2}{3}]\right)}{
 4 \sqrt[6]{3}\sqrt{-\sqrt{3} c_3 + 3 c_4} \sqrt{\Gamma[\frac{1}{3}]}}}, \]
 which can be integrated to give,
 \[r_1=\frac{2\sqrt{-\sqrt{3} c_3 + 3 c_4}\sqrt{2\Gamma[\frac{1}{3}]}}{\sqrt{\sqrt{3} c_3 + 3 c_4}
 \Gamma[\frac{2}{3}]}\ln\left\{\frac{\left[- 2 \sqrt[6]{3}c_3 \Gamma\left[\frac{1}{3}\right]+  2c_4\sqrt[3]{9}
 \Gamma\left[\frac{1}{3}\right]+t_0\left(\left[\sqrt{3} c_3 + 3 c_4\right]\Gamma\left[\frac{2}{3}\right]\right) \right]}
 {\left[- 2 \sqrt[6]{3}c_3 \Gamma\left[\frac{1}{3}\right]+  2c_4\sqrt[3]{9}\Gamma\left[\frac{1}{3}\right]+t_1\left(\left[\sqrt{3} c_3 + 3 c_4\right]
 \Gamma\left[\frac{2}{3}\right]\right) \right]}\right\},\]
 \[\Rightarrow r_1=\frac{2\sqrt{-\sqrt{3} c_3 + 3 c_4}\sqrt{2\Gamma[\frac{1}{3}]}}{\sqrt{\sqrt{3} c_3 + 3 c_4}
 \Gamma[\frac{2}{3}]}\ln\left\{1+Z\right\},\]
where $\Gamma$ is the gamma function.  Then from (26) we have,\be
s=R_0r_1=\frac{\ln\left[- 2 \sqrt[6]{3}c_3
\Gamma\left[\frac{1}{3}\right]+
2c_4\sqrt[3]{9}\Gamma\left[\frac{1}{3}\right]+t_0\left(\left[\sqrt{3}
c_3 + 3 c_4\right]\Gamma\left[\frac{2}{3}\right]\right)
\right]}{\sqrt{\sqrt{3} c_3 + 3 c_4}\Gamma[\frac{2}{3}]}
\ln\left\{1+Z\right\}.\ee From Eq. (27) we see that  there is no
particle horizons since when $Z\rightarrow \infty$ the proper
distance $s$ tends to infinity as well.

 Using the same method we can show that radiation and stiff matter models  $(\omega=\frac{4}{3}, k=0, \Lambda(t)=\epsilon_0+\epsilon_1t)$  and $(\omega=2, k=0, \Lambda(t)=\epsilon_0+\epsilon_1t)$, respectively,  are free from particle horizons.
\newsection{Main results and discussion}
In the present work, we consider cosmological  application  in the domain of KK-theory, taking  the Newtonian  and the cosmological constants,
$G$ and $\Lambda$,  to be functions of time $t$. Assuming  homogeneity and isotropy, the field equations of KK-theory give two differential equations.
 We have added to them an equation of sate and two further conditions related to conservation. This is done in order to get the five unknown functions
 $R(t)$, $P(t)$, $\rho(t)$, $\Lambda(t)$ and $G(t)$. One of these equations shows, in a clear way, that the relation  between $\Lambda(t)$ and $G(t)$
 indicates  that $\Lambda(t)$ is constant whenever $G(t)$ is constant and vice versa. We study the flat space, i.e., $k=0$, and
  assume a Taylor expansion of $\Lambda(t)$ keeping up to first  order of t i.e.,
  $O\left(t\right)$. We can see from Eqs. (12), (17), (19) and (21) that
  the values of  $\Lambda(t)$ and the scale factor $R(t)$  are finite  while, when  $t\rightarrow
  0$  the values of  $\Lambda(t)$ and the scale factor $R(t)$ are unbounded when $t\rightarrow
  \infty$.

   A summary of the results obtained  are compared below in the following  table.
   Different figures  related to this summary are given. From these figures we can conclude:\vspace{.3cm}\\
i)  The models resulting from the  three studied cases, i.e., dust, radiation and stiff matter have no singularity
unless when $A_0=\left(\frac{{\epsilon_1}}{12}\right)^{1/3}$, $B_0=\left(\frac{{\epsilon_1}^3}{16}\right)^{1/3}$
and ${C^\ast}_0=\left(\frac{{\epsilon_1}}{24}\right)^{1/3}$ vanish,  which contradicts our assumption that $\Lambda(t)=\epsilon_0+\epsilon_1 t$
 is a function of $t$. In other words the models are singular when $\Lambda(t)$ is constant.\vspace{.3cm}\\
ii) The  resulting models  are accelerating which is consistent with  most recent observation$^{ \rm [1,2,3]}$.\vspace{.3cm}\\
iii) Figure 3 shows that density in the three models are decreasing with time which  ensures that our universe is expand. \vspace{.3cm}\\
iv) The  three models are free from particle horizons, which is  consistent the previous result$^{ \rm [53]}$\vspace{.3cm}\\
v) We have considered the cosmological term to contribute to the material-energy contents of the universe.\vspace{.3cm}\\
vi) Eq. (7) satisfy the conservation law when $G(t)$ and
$\Lambda(t)$ are function of time in contradiction with the result
obtained before
(Eq. (8) \cite{Sm1}).\vspace{.3cm}\\
\begin{center}
Table (I)  comparison between the results of scale factors $R(t)$
Hubble parameters $H(t)$ density $\rho(t)$ and deceleration
parameters $q(t)$ for $\Lambda(t)=\epsilon_0+\epsilon_1
t$\footnote{$f(t)=\frac{\Lambda(t)}{A_{00}}$,
$A_{00}=\left(\frac{3{\epsilon_1}^2}{2}\right)^{1/3}$,
$f'(t)=\left(1,\frac{\Lambda(t)}{A_{00}}\right)$,
$f_1(t)=\frac{\Lambda(t)}{B_{00}}$,
$B_{00}=\left(\frac{27{\epsilon_1}^2}{32}\right)^{1/3}$,
$f_1'(t)=\left(1,\frac{\Lambda(t)}{B_{00}}\right)$
$f_2(t)=\frac{\Lambda(t)}{C_{00}}$, $
C_{00}=\left(\frac{3{\epsilon_1}^2}{8}\right)^{1/3}$ and
$f_2'(t)=\left(1,\frac{\Lambda(t)}{C_{00}}\right)$.}
\end{center}

\begin{center}
{\tiny
\begin{tabular}{|l|l|l|l|}

\hline
&&&\\
Equation  &   \hspace{2cm}$\omega=1$  & \hspace{2cm}
$\omega=4/3$&\hspace{2cm} $\omega=2$ \\ of state & & & \\ \hline
&&&\\
 Scale Factor &$\sqrt{\frac{c_3 A(f(t))-c_4 B(f(t))}{A_0\left[A(f'(t))
 B\left(f(t)\right)-B(f'(t))
 A\left(f(t)\right)\right]}}$  &$\left\{ \frac{c_5 A(f_1(t))-c_6B(f_1(t))}{B_0\left[A(f'_1(t))
 B(f_1(t))-B(f'_1(t))
 A(f_1(t))\right]}\right\}^{\frac{3}{8}}$&$\left\{\frac{c_7 A(f_2(t))-c_8 B(f_2(t))}{C_{0}\left[A(f'_2(t))
 B(f_2(t))-B(f'_2(t))
 A(f_2(t))\right]}\right\}^{1/4}$\\
 &&&\\\hline
 &&&\\
  Hubble &$ A_0\frac{c_3 A(f'(t))-c_4B(f'(t))}{c_3 A(f(t))-c_4B(f(t))}$  &$B_0\frac{c_5 A(f'_1(t))c_6 B(f'_1(t))}{c_5 A(f_1(t))-c_6 B(f_1(t))}$& $C_0\frac{c_7 A(f'_2(t))-c_8 B(f'_2(t))}{c_5 A(f_2(t))+B(f_2(t))}$ \\ parameter & & & \\ \hline
  &&&\\
  Density &$\frac{c_1 \left(A_0\left[A(f'(t)) B(f(t))-B(f'(t))
 A(f(t))\right]\right)^2}{\left(c_3 A(f(t))-c_4B(f(t))\right)^2}$ &$c_1\left\{ \frac{B_0\left[A(f'_1(t)) B(f_1(t))-B(f'_1(t))
 A(f_1(t))\right]}{c_5 A(f_1(t))-c_6B(f_1(t))}\right\}^{\frac{3}{2}}$&$\frac{c_1C_{0}\left[A(f'_2(t))
 B(f_2(t))-B(f'_2(t))
 A(f_2(t))\right]}{c_7 A(f_2(t))-c_8 B(f_2(t))}$ \\ &&& \\\hline
 &&&\\
  Deceleration  &$\displaystyle\frac{1}{3\sqrt[3]{{\epsilon_1}^2}\left[c_3A(f'(t)) -c_4B(f'(t))\right]^2}$  & $\displaystyle\frac{1}{18\sqrt[3]{{4\epsilon_1}^2}\left[c_5A(f_1'(t)) -c_6B(f_1'(t))\right]^2}$ &$\displaystyle\frac{1}{3\sqrt[3]{{\epsilon_1}^2}\left[c_7A(f_2'(t)) -c_8B(f_2'(t))\right]^2}$\\
 Parameter  &$\Biggl\{3\sqrt[3]{{\epsilon_1}^2} \left[c_3A(f'(t)) -c_4B(f'(t))\right]^2$ &$\Biggl\{15\sqrt[3]{{2\epsilon_1}^2} \left[c_5A(f_1'(t)) -c_6B(f_1'(t))\right]^2$ &$\Biggl\{3\sqrt[3]{{3\epsilon_1}^2} \left[c_7A(f_2'(t)) -c_8B(f_2'(t))\right]^2$ \\
  & $-\sqrt[3]{144}\Lambda(t)\left[c_3A(f(t)) -c_4B(f(t))\right]^2 \Biggr\}$ & $-32\Lambda(t)\left[c_5A(f_1(t)) -c_6B(f_1(t))\right]^2 \Biggr\}$  &$-8\Lambda(t)\left[c_7A(f_2(t)) -c_8B(f_2(t))\right]^2 \Biggr\}$ \\ &&& \\\hline
  &&&\\
 Particle  &$\displaystyle\frac{ \ln\left\{1+Z\right\}}{\sqrt{\sqrt{3} c_3 + 3 c_4}\Gamma[\frac{2}{3}]}$  & $\displaystyle\frac{ \ln\left\{1+Z\right\}}{\sqrt{\sqrt{3} c_5 + 3 c_6}\Gamma[\frac{2}{3}]}$ &$\displaystyle\frac{ \ln\left\{1+Z\right\}}{\sqrt{\sqrt{3} c_7 + 3 c_8}\Gamma[\frac{2}{3}]}$\\
Horizons  &$\Biggl\{\ln\Biggl[- 2 \sqrt[6]{3}c_3 \Gamma\left[\frac{1}{3}\right]+  2c_4\sqrt[3]{9}\Gamma\left[\frac{1}{3}\right]$ &$\Biggl\{\ln\Biggl[- 2 \sqrt[6]{3}c_5 \Gamma\left[\frac{1}{3}\right]+  2c_6\sqrt[3]{9}\Gamma\left[\frac{1}{3}\right]$ &$\Biggl\{\ln\Biggl[- 2 \sqrt[6]{3}c_7 \Gamma\left[\frac{1}{3}\right]+  2c_8\sqrt[3]{9}\Gamma\left[\frac{1}{3}\right]$ \\
  & $+t_0\left(\left[\sqrt{3} c_3 + 3 c_4\right]\Gamma\left[\frac{2}{3}\right]\right) \Biggr] \Biggr\}$ & $+t_0\left(\left[\sqrt{3} c_5 + 3 c_6\right]\Gamma\left[\frac{2}{3}\right]\right) \Biggr] \Biggr\}$  &$+t_0\left(\left[\sqrt{3} c_7 + 3 c_8\right]\Gamma\left[\frac{2}{3}\right]\right) \Biggr] \Biggr\}$ \\ &&& \\\hline
\end{tabular}
 }
\end{center}

\end{document}